\documentclass[prb,preprint,showpacs,amsmath,amssymb,superscriptaddress]{revtex4}
\usepackage{epsfig}
\usepackage{graphicx}
\usepackage{amsmath,amssymb}

\newcommand{\beq}[1]{
\begin{equation}
\label{e#1} }

\newcommand{\eeq}{
\end{equation}
}

\begin{document}

\title{Antiferromagnetic I-Mn-V semiconductors}

\author{T.~Jungwirth}
\affiliation{Institute of Physics ASCR, v.v.i., Cukrovarnick\'a 10, 162 53 Praha 6, Czech Republic}
\affiliation{School of Physics and Astronomy, University of Nottingham, Nottingham NG7 2RD, United Kingdom}
\author{V. Nov\'ak}
\affiliation{Institute of Physics ASCR, v.v.i., Cukrovarnick\'a 10, 162 53 Praha 6, Czech Republic}
\author{X. Mart\'{i}}
\affiliation{Faculty of Mathematics and Physics, Charles University in Prague, Ke Karlovu 3, 121 16 Prague 2, Czech Republic}
\author{M. Cukr}
\affiliation{Institute of Physics ASCR, v.v.i., Cukrovarnick\'a 10, 162 53 Praha 6, Czech Republic}
\author{F.~M\'aca}
\author{A.~B.~Shick}
\author{J.~Ma\v{s}ek}
\affiliation{Institute of Physics ASCR, v.v.i., Na Slovance 2, 182 21 Praha 8, Czech Republic}
\author{P.~Horodysk\'a}
\affiliation{Faculty of Mathematics and Physics, Charles University in Prague, Ke Karlovu 3, 121 16 Prague 2, Czech Republic}
\author{P.~N\v{e}mec}
\author{V.~Hol\'{y}}
\affiliation{Faculty of Mathematics and Physics, Charles University in Prague, Ke Karlovu 3, 121 16 Prague 2, Czech Republic}
\author{J. Zemek}
\affiliation{Institute of Physics ASCR, v.v.i., Cukrovarnick\'a 10, 162 53 Praha 6, Czech Republic}
\author{P.~Ku\v{z}el}
\affiliation{Institute of Physics ASCR, v.v.i., Na Slovance 2, 182 21 Praha 8, Czech Republic}
\author{I.~N\v{e}mec}
\affiliation{Faculty of Science, Charles University in Prague, Hlavova 2030, 128 40 Prague 2, Czech Republic}
\author{B.~L.~Gallagher}
\author{R.~P.~Campion}
\author{C.~T.~Foxon}
\affiliation{School of Physics and Astronomy, University of Nottingham, Nottingham NG7 2RD, United Kingdom}
\author{J.~Wunderlich}
\affiliation{Institute of Physics ASCR, v.v.i., Cukrovarnick\'a 10, 162 53 Praha 6, Czech Republic}
\affiliation{Hitachi Cambridge Laboratory, Cambridge CB3 0HE, United Kingdom}

\date{\today}

\maketitle

{\bf After decades of research, the low Curie temperature of ferromagnetic semiconductors remains the key problem in the development of magnetic-semiconductor spintronic technologies. Removing this roadblock might require a change of the field's basic materials paradigm by looking  beyond ferromagnets. Recent studies of relativistic magnetic and magneto-transport anisotropy effects, which in principle are equally well present in materials with ferromagnetically and antiferromagnetically ordered spins,  have inspired our search for antiferromagnetic semiconductors suitable for high-temperature spintronics. Since these are not found among the magnetic counterparts of common III-V or II-VI semiconductors, we turn the attention in this paper to high N\'eel temperature I-II-V magnetic compounds whose electronic structure has not been previously identified. Our combined experimental and theoretical work on LiMnAs provides basic prerequisite for the systematic research of this class of materials by demonstrating the feasibility to grow single crystals of group-I alkali metal compounds by molecular beam epitaxy, by demonstrating the semiconducting  band structure of the I-Mn-V's, and by analyzing their spin-orbit coupling characteristics favorable for spintronics.}


All current spintronic applications are based on ferromagnetically (FM) ordered spins of transition metals such as Ni, Co, Fe and their alloys.\cite{Chappert:2007_a} These material systems have provided  the strong magneto-resistive effects essential to the development of  commercial spintronics. However, there are fundamental physical limitations for FM metal materials which may make them impractical to realize the full potential of spintronics. Metals are unsuitable for transistor and information processing applications, for opto-electronics, and the large magnetic stray fields produced by these materials also make them unfavorable for high-density integration. A considerable effort has been directed towards overcoming these limitations by developing FM semiconductor materials based on conventional semiconductor hosts.  (III,Mn)V compounds are among the most extensively explored examples of these systems.\cite{Dietl:2008_b} They do not allow for the high-temperature operation but they are ideal test bed materials for exploring new spintronic concepts. 

The  inspiration for the work presented in this paper comes from discoveries of relativistic magnetic and magneto-transport anisotropy effects in (III,Mn)V nano-devices,\cite{Wunderlich:2006_a,Dietl:2008_b} which have the desired large magnitudes  and whose common characteristics is that they are an even function of the microscopic magnetic moment vector. The concept, whose generic validity has been recently confirmed in transition metal FMs,\cite{Gao:2007_a, Park:2008_a,Bernand-Mantel:2009_a} paves a way to spintronics in a wide range of systems beyond FMs, including materials with antiferromagnetically (AFM) coupled moments.\cite{Shick:2010_a} When realized in AFM semiconductors, these relativistic spintronic effects can be combined with electric-field gating in spintronic transistor structures. Applications of AFM semiconductors may also include  integration of  conventional semiconductor micro and opto-electronics functionalities directly  in the exchange-biasing AFM layers in common spintronic devices. The aim of this paper is to identify a suitable class of materials for high-temperature AFM semiconductor spintronics.  

In Fig.~1(a) we illustrate that  a strong AFM ground state is from the basic physics perspective much more compatible with a gapped, semiconductor-like band structure than a strong FM state. In the FM case, the gap competes with the exchange spin-splitting of the bands which at strong FM coupling turns the system into a metal. In AFMs, this competition is missing and indeed a large majority of magnetic semiconductors order antiferromagnetically. 
Our interest is in magnetic compounds which are direct counterparts of the most common non-magnetic semiconductors with eight valence electrons per primitive cell. Fig.~1(b) illustrates a survey of these materials and the search path that has brought our attention to the I-Mn-V compounds. The common non-magnetic semiconductors are derived from the group-IV (Si, Ge) semiconductors by applying the "proton transfer" rule:\cite{Harrison:1980_a} By imagining a transfer of a proton from one to the other group-IV atom in the primitive cell we obtain the III-V compounds, a transfer of two protons gives the II-VI materials, etc. The magnetic counterpart to the group-II atoms with a half-filled $d$($f$) shell is Mn 3$d^5$4$s^2$ (Eu 4$f^7$6$s^2$) and the neighboring magnetic atom Fe (Gd)  can be considered as a group-III element when searching for the magnetic semiconductors. 

Since there is no direct 
realization of a magnetic semiconductor within the  group-IV crystals we need to proceed to compounds. Here FeAs is an example of AFM semiconductors from the III-V family. As implied by Fig.~1(a), the less frequent FM semiconductors have a better chance to occur among the more localized (less hybridized) $f$-electron magnetic elements and the lighter anions (wider gap). GdN FM semiconductor belongs to this group of materials. Transition temperatures of all the III-V magnetic semiconductors are  below room-temperature and the non half-filled shell of the magnetic element is one of the reasons for weaker magnetic interactions. Indeed, (Ga,Mn)As already suggests that using Mn with the half-filled magnetic shell, i.e. highest moment among the 3$d$  magnetic elements, has a favorable effect on the strength of  magnetic coupling. (Recall that the (Ga,Mn)As random alloy is not an intrinsic but rather a heavily doped degenerate semiconductor because Mn$_{\rm Ga}$ is not an isovalent impurity.) Consistent with this trend there is  a number of Mn-chalcogenide AFM semiconductors and there are also several FMs among the $f$-electron and lighter anion II-VI compounds. Still the transition temperatures of II-VI magnetic semiconductors do not safely exceed room temperature. Here the larger anions and more ionic bonds compared to the III-V's result in larger lattice parameters which softens the magnetic interactions. The idea behind our work discussed in the following paragraphs  is that by transferring only one proton to the anion and the other proton to a new lattice site should result in  I-Mn-V semiconductors which combine the favorable tighter lattice arrangement of the group-V pnictides and the large magnetic moment on Mn. 

Alkali-metal I-Mn-V compounds are stable materials which have been previously prepared by chemical synthesis in polycrystalline or powder forms. X-ray studies  showed that  the I-Mn-V's, as well as the non-magnetic I-II-V's, have a similar crystal structure to III-V semiconductors, as illustrated in Fig.1(c).\cite{Bacewicz:1988_a,Achenbach:1981_a,Bronger:1986_a,Muller:1991_a,Schucht:1999_a}  Neutron-diffraction measurements  identified AFM coupling in the Mn-planes persisting to very high temperatures and a weaker interlayer coupling along the c-axis which persists in most of these compounds above 400~K.\cite{Bronger:1986_a,Muller:1991_a,Schucht:1999_a} While the chemists have provided us with the knowledge of these  favorable structural and magnetic characteristics of I-Mn-V's, the compounds have  been virtually unexplored by the physical materials research community. In particular there is no preexisting information on the electronic structure of I-Mn-V's and on the synthesis of alkali-metal group-I compounds  by modern molecular beam epitaxy (MBE) growth techniques. We now proceed to address these two basic problems in the materials science of I-Mn-V compounds.

For the epitaxial growth of LiMnAs we used InAs substrate because the crystallographically equivalent As planes in InAs and in LiMnAs have very similar lattice parameters (4.283~\AA~in InAs versus 4.267~\AA~in LiMnAs), as inferred from X-ray data on the LiMnAs materials prepared previously by chemical synthesis. The comparison between the respective crystal structures (see Fig. 1(c)) further implies that starting from the common As plane, LiMnAs should grow epitaxially under a tensile strain with its Mn planes rotated by 45$^{\circ}$ with respect to the counterpart In planes of the substrate. Since MnAs is a ferromagnetic metal with a hexagonal crystal structure we have also prepared reference MnAs samples deposited on InAs to highlight the striking consequences of Li incorporation during the growth.

For this work we have grown ten LiMnAs/InAs wafers in a Kryovak MBE system equipped with Li, Mn, and As solid sources. Before the growth, the surface oxide on the InAs substrate was desorbed in the As atmosphere at 450$^{\circ}$C. The LiMnAs films were deposited directly on the substrate at low temperatures ($\sim150^{\circ}$C) without growing any preceding buffer layer. The respective cell temperatures were 430$^{\circ}$C (Li), 840$^{\circ}$C (Mn), and 260$^{\circ}$C (As), producing fluxes of the elements at approximately stoichiometric ratio 1:1:1. The reproducibility of the growth conditions have been tested over many growth cycles, showing no degradation or apparent damage to the MBE system  due to the presence of large amounts of Li.

The reflection high energy electron diffraction (RHEED) images were recorded {\em in situ} during the growth. Within the first few minutes of the growth a sharp and stable 1$\times$1 RHEED pattern, shown in Fig.~2(a), emerged from the original pattern of the $c(4\times4)$ reconstructed surface of the InAs substrate. The LiMnAs RHEED images  demonstrate the 2D growth mode of a high quality epilayer and confirm the expected in-plane cubic symmetry of the LiMnAs crystal. The results are in striking contrast to the measurement of the reference MnAs epilayer, shown in Fig.~2(b), deposited on the InAs substrate under the same growth conditions, except for the Li cell remaining closed during the growth. The MnAs sample shows asymmetric and low epitaxial quality RHEED pattern as a result of the large lattice mismatch of MnAs.

In parallel to the RHEED, the growth process of the LiMnAs epilayers was monitored by  {\em in situ} measurements of Fabry-P\'erot interference oscillations, shown in Fig.~2(c). In the experiment, the light emitted by the cells in the MBE system, reflected by the growing film, and transmitted outside the chamber through an optical port was recorded  by a spectrometer in a spectral range of 870 to 1400 nm (0.855 to 1.425~eV). The presence of the Fabry-P\'erot oscillations up to large film thickness is a strong  indication of the semiconducting nature of LiMnAs. Note, that the LiMnAs film thicknesses, and the derived growth rate of approximately 200 nm per hour, were obtained by measuring the final thickness profile of the wafers across the edges masked by the sample holder during the growth, as shown in Fig.~2(d). The 20\% uncertainty in the determination of the LiMnAs film thickness is due to the surface oxide layer which forms in air and/or due to the capping layer deposited onto the LiMnAs film to prevent its oxidation. Fabry-P\'erot  oscillations in the upper part of Fig.~2(c) correspond to the InAs cap and are consistent with its 3.5 refractive index. The lower part recorded during the LiMnAs growth yields substantially smaller refractive index (by a factor of 1.5--2) suggesting that LiMnAs has a larger band gap than InAs. 
The incorporation of Li and Mn in the grown epilayers was confirmed {\em ex situ} by a series of sputtering and X-ray photoemission measurements which showed, within the experimental scatter,  the expected 1:1 ratio of Li and Mn inside the film. 
X-ray diffraction experiments presented in Fig.~3 prove that the films are epitaxial and single phase LiMnAs crystals. In addition to the InAs substrate peaks, the diffraction curve in Fig.~3(a) shows  the full set of (001) oriented LiMnAs reflections and no traces of other phases or orientations. Note that the peaks are slightly shifted to higher angles with respect to the bulk values (denoted by crosses) in agreement with the expected tensile strain in the LiMnAs epilayer on InAs. In order to investigate the epitaxial relationship, we performed azimuthal scans as a function of the wavevector $Q$ (see Fig.~3(b)). The data show that the LiMnAs(102) and InAs(131) reflections are separated by the angle $\phi = 26.7^{\circ}$ which matches the nominal separation  in the case of a 45$^{\circ}$ in-plane rotation of the LiMnAs unit cell with respect to the substrate. Hence, the [110]LiMnAs direction is parallel to the [100]InAs direction, as illustrated in the schematic 3D diagram in Fig.~3(b). Reciprocal space maps shown in Fig.~3(c) evidence the vertical alignment of the substrate and film peak, i.e.,  the LiMnAs film is an epitaxial single-crystal fully strained to the InAs substrate. The black and red crosses in the plot denote the expected positions for the substrate and bulk LiMnAs, respectively. Due to the in-plane tensile strain, the out-of-plane parameter is contracted leading to the displacement of the LiMnAs(204) reflection. As a result of this small structural distortion, the unit cell volume of the LiMnAs epilayer is increased by 0.2\%. 

Based on the equivalence between the lattice structure of our LiMnAs single-crystals and of the previously chemically synthesized polycrystalline bulk materials we expect the same AFM structure of the epilayers, as illustrated in Fig.~1(c). Neutron diffraction measurements on thin film epilayers are not routinely available and attempts to perform these magnetic structure experiments on LiMnAs are beyond the scope of this initial work. Instead we performed superconducting quantum interference device  (SQUID) measurements of the magnetization which are shown in Fig.~3(d). The data are consistent with the picture of compensated Mn moments in the AFM LiMnAs. The comparison of the temperature-dependent remanence in LiMnAs and in the reference FM MnAs with the same amount of Mn in the epilayer, as well as the comparison of the low-temperature saturation in LiMnAs, MnAs, and of the theoretical Brillouin function of uncoupled Mn $S=5/2$ moments, rule out ferromagnetic and paramagnetic behavior of Mn in our LiMnAs epilayers. In agreement with the high N\'eel temperature of the chemically synthesized bulk materials we found no signatures of the vicinity of the critical point in the susceptibility in the explored temperature range up to 400~K.

The {\em ex situ} optical transmission measurements of the LiMnAs grown on InAs is shown in Fig.~3(e).  The observed transparency of the wafer (up to the band-gap energy of the 0.5~mm thick InAs substrate) complements the above {\em in situ} optical demonstration of the semiconducting character of LiMnAs. The optical transparency of LiMnAs is in  striking contrast to the control MnAs sample which, due to its metallic band structure, is strongly absorbing over the entire studied spectral range, as shown in Fig.~3(e). We note that the enhanced absorption in LiMnAs/InAs wafer at the low-energy side of the spectrum is due to free carriers introduced by Li diffused into the InAs substrate during the growth. Interstitial Li acts as a shallow donor in InAs and the resulting n-type doping of InAs can be as high as $\sim 10^{18}-10^{19}$~cm$^{-3}$ under our growth conditions. We made these observations based on our control experiment in which the InAs substrate was exposed in the MBE chamber to the Li flux alone. In the resulting InAs:Li we observe again the enhanced low-energy absorption which is correlated with high dc conductivity of the sample. After annealing the Li out of the InAs, the transparency at low energies is recovered and the conductivity drops to the nominal value of the unprocessed substrate. 

We now proceed to the theoretical investigation of the electronic structure of the I-Mn-V compounds. We have performed band structure calculations of LiMnAs, NaMnAs, and KMnAs  using full-potential density-functional theory. We found that the AFM state has always lower energy than the FM state. In LiMnAs, the difference in GGA is 34.4~mRy/atom (32.7~mRy/atom in LDA) and similarly in NaMnAs and KMnAs the AFM state is lower in energy by 32.9 and 32.5~mRy/atom, resp. Remarkably, these values, which can be used to  estimate  the  N\'eel temperature $T_N$, are larger in the I-Mn-V's than in metal Mn-based alloys whose $T_N\sim 10^3$~K. \cite{Khmelevskyi:2008_a} Our calculations therefore reproduce not only the ground state AFM structure of I-Mn-V's but also explain the  high $T_N$.

The I-Mn-V AFMs are intrinsic semiconductors for the stoichiometric composition 1:1:1 of the constituent elements, as shown by LDA band dispersions in Fig.~4(a)-(d). Note, that the calculated band gaps  are likely underestimated by the density functional theory. Besides the common shortcomings of the LDA, the accurate theoretical determination of the band gap is further complicated by correlation effects on Mn $d$-orbitals. The states are on one hand strongly admixed  in the top valence band and bottom conduction band states, as shown by the density of states (DOS) plotted in Fig.~5(a),(b), but on the other hand are still significantly more localized then the $sp$-states. Accounting for  these effects in LiMnAs  increases significantly the band gap and suppresses the relative difference between the indirect and direct gaps. In KMnAs, the correlations even change the character of the material from an indirect to a direct gap semiconductor.  

The strong admixture of Mn $d$-orbitals in the valence and conduction band DOSs and correlations on these states have a significant effect on the dielectric properties of I-Mn-V's. As shown in Fig.~5(c), the increase of the band gap in LiMnAs in the LDA+U and the corresponding shift of the absorption edge in the imaginary part of the dielectric function $\epsilon$ are correlated with the suppression of the real part of $\epsilon$. The large suppression by nearly a factor of 2 is partly due to the common scaling in semiconductors between band gap and the refractive index but, to a large extent, also due to the optical transitions involving the $d$-orbitals. These transitions contribute strongly to the dielectric function and their shift to higher energies in the LDA+U further reduces the real part of $\epsilon$ and of the refractive index. The resulting values are consistent with the low refractive index of LiMnAs inferred from the Fabry-P\'erot oscillations in Fig.~2(c). Similarly, the significantly larger band gap of LiMnAs compared to InAs is consistent with optical measurements on our LiMnAs/InAs epilayers.

The magneto-crystalline anisotropy energy (MAE) is an important example of phenomena based on the collective exchange interaction and spin-orbit coupling which are even functions of the microscopic moment vector and are therefore present not only in ferromagnets but also in compensated AFMs. Our relativistic full-potential LDA calculations show that spin-orbit coupling on both the magnetic Mn and on the group-V elements contribute to the large net MAE. The uniaxial in-plane vs out-of-plane anisotropy constant in LiMnAs is as high as 0.58~meV per formula unit. The MAE is one of the key parameters determining the interface coupling between a FM and an AFM which is utilized, e.g., for pinning the FM via the exchange bias effect in giant  or tunneling magneto-resistance  spintronic sensor and memory devices.\cite{Chappert:2007_a} The coupling can be also used for controlling the staggered moment orientation in the AFM by the exchange spring effect induced by rotating moments in the ferromagnet.\cite{Scholl:2004_a} In an AFM alone, the MAE  can be used to rotate the staggered moments by applying piezoelectric or electrostatic fields.\cite{Shick:2010_a} Note that the electrostatic effects are subtle and conceptually difficult to describe in metallic systems due to the strong screening effects of high-density free carriers.  The intrinsic or weakly doped semiconductor character of I-Mn-V AFMs makes them particularly suitable and unique systems for achieving large electrostatic gating effects on the MAE.  

Similar to the MAE, there is a class of  anisotropic  magneto-transport phenomena in bulk and nanostructured magnetic materials which are an even function of the microscopic  moment. In very general terms, the physical origin of these effects is the anisotropy in the DOS with respect to the  orientation of magnetic moments. Our DOS calculations shown in Fig.~5(d) imply that these effects can be order of magnitude larger in the I-Mn-V AFMs compared to metal Mn-based AFMs\cite{Shick:2010_a} and can be tuned by varying the carrier concentration via doping or electrical gating in field-effect transistor structures. We note that no signficant qualitative differences were found when comparing the magnetic anisotropy effects in LiMnAs calcuated in the LDA and LDA+U.


To conclude, we have presented a new class of semiconductor materials which opens the prospect for high-temperature semiconductor spintronics. We have made an observation that the studied I-Mn-V AFM compounds are the simplest magnetic counterparts to conventional eight valence electron semiconductors which offer this prospect. In the experimental part of our study we have demonstrated on LiMnAs that high-quality single-crystals of group-I compounds can be grown by molecular beam epitaxy. Our {\em ab initio} calculations confirmed that the I-Mn-V compounds are strong AFMs, confirmed our prediction and experimental indications of the semiconducting band structure of these materials, and unveiled strong spin-orbit coupling character of conduction and valence band states favorable for spintronics. These experimental and theoretical results define a framework for future systematic research of materials properties of the broad family of I-Mn-V compounds and of their utility in semiconductor nanostructures and spintronic devices.

\bibliographystyle{naturemag}

\begin{thebibliography}{10}
\expandafter\ifx\csname url\endcsname\relax
  \def\url#1{\texttt{#1}}\fi
\expandafter\ifx\csname urlprefix\endcsname\relax\def\urlprefix{URL }\fi
\providecommand{\bibinfo}[2]{#2}
\providecommand{\eprint}[2][]{\url{#2}}

\bibitem{Chappert:2007_a}
\bibinfo{author}{Chappert, C.}, \bibinfo{author}{Fert, A.} \&
  \bibinfo{author}{Dau, F. N.~V.}
\newblock \bibinfo{title}{The emergence of spin electronics in data storage}.
\newblock \emph{\bibinfo{journal}{Nature Materials}}
  \textbf{\bibinfo{volume}{6}}, \bibinfo{pages}{813} (\bibinfo{year}{2007}).

\bibitem{Dietl:2008_b}
\bibinfo{author}{eds. T.~Dietl}, \bibinfo{author}{Awschalom, D.~D.},
  \bibinfo{author}{Kaminska, M.} \& \bibinfo{author}{Ohmo, H.}
\newblock \bibinfo{title}{Spintronics}.
\newblock In \emph{\bibinfo{booktitle}{Spintronics}}, vol.~\bibinfo{volume}{82}
  of \emph{\bibinfo{series}{Semiconductors and Semimetals}}
  (\bibinfo{publisher}{Elsevier}, \bibinfo{year}{2008}).

\bibitem{Wunderlich:2006_a}
\bibinfo{author}{Wunderlich, J.} \emph{et~al.}
\newblock \bibinfo{title}{Coulomb blockade anisotropic magnetoresistance effect
  in a (ga,mn)as single-electron transistor}.
\newblock \emph{\bibinfo{journal}{Phys. Rev. Lett.}}
  \textbf{\bibinfo{volume}{97}}, \bibinfo{pages}{077201}
  (\bibinfo{year}{2006}).

\bibitem{Gao:2007_a}
\bibinfo{author}{Gao, L.} \emph{et~al.}
\newblock \bibinfo{title}{Bias voltage dependence of tunneling anisotropic
  magnetoresistance in magnetic tunnel junctions with mgo and al2o3 tunnel
  barriers}.
\newblock \emph{\bibinfo{journal}{Phys. Rev. Lett.}}
  \textbf{\bibinfo{volume}{99}}, \bibinfo{pages}{226602}
  (\bibinfo{year}{2007}).

\bibitem{Park:2008_a}
\bibinfo{author}{Park, B.~G.} \emph{et~al.}
\newblock \bibinfo{title}{Tunneling anisotropic magnetoresistance in
  {multilayer-(Co/Pt)/AlO$_{x}$/Pt} structures}.
\newblock \emph{\bibinfo{journal}{Phys. Rev. Lett.}}
  \textbf{\bibinfo{volume}{100}}, \bibinfo{pages}{087204}
  (\bibinfo{year}{2008}).

\bibitem{Bernand-Mantel:2009_a}
\bibinfo{author}{Bernand-Mantel, A.} \emph{et~al.}
\newblock \bibinfo{title}{Anisotropic magneto-coulomb effects and magnetic
  single-electron-transistor action in a single nanoparticle}.
\newblock \emph{\bibinfo{journal}{Nat. Phys.}} \textbf{\bibinfo{volume}{5}},
  \bibinfo{pages}{920} (\bibinfo{year}{2009}).

\bibitem{Shick:2010_a}
\bibinfo{author}{Shick, A.~B.}, \bibinfo{author}{Khmelevskyi, S.},
  \bibinfo{author}{Mryasov, O.~N.}, \bibinfo{author}{Wunderlich, J.} \&
  \bibinfo{author}{Jungwirth, T.}
\newblock \bibinfo{title}{Spin-orbit coupling induced anisotropy effects in
  bimetallic antiferromagnets: A route towards antiferromagnetic spintronics}.
\newblock \emph{\bibinfo{journal}{Phys. Rev. B}} \bibinfo{pages}{in press}
  (\bibinfo{year}{2010}).

\bibitem{Harrison:1980_a}
\bibinfo{author}{Harrison, W.~A.}
\newblock \emph{\bibinfo{title}{Electronic Structure and the Properties of
  Solids}} (\bibinfo{publisher}{Freeman, San Francisco}, \bibinfo{year}{1980}).

\bibitem{Bacewicz:1988_a}
\bibinfo{author}{Bacewicz, R.} \& \bibinfo{author}{Ciszek, T.~F.}
\newblock \bibinfo{title}{Preparation and characterization of some aibiicv type
  semiconductors}.
\newblock \emph{\bibinfo{journal}{Appl. Phys. Lett.}}
  \textbf{\bibinfo{volume}{52}}, \bibinfo{pages}{1150} (\bibinfo{year}{1988}).

\bibitem{Achenbach:1981_a}
\bibinfo{author}{Achenbach, G.} \& \bibinfo{author}{Schuster, H.~U.}
\newblock \bibinfo{title}{Ternary compounds of lithium and sodium with
  manganese and elements of the fifth main group}.
\newblock \emph{\bibinfo{journal}{Z. anorg. allg. Chem.}}
  \textbf{\bibinfo{volume}{475}}, \bibinfo{pages}{9} (\bibinfo{year}{1981}).

\bibitem{Bronger:1986_a}
\bibinfo{author}{Bronger, W.}, \bibinfo{author}{{M\"{u}ller}, P.},
  \bibinfo{author}{{H\"{o}ppner}, R.} \& \bibinfo{author}{Schuster, H.~U.}
\newblock \bibinfo{title}{The magnetic properties of NaMnP, NaMnAs, NaMnSb,
  NaMnBi, LiMnAs, and KMnAs, characterized by neutron diffraction experiments}.
\newblock \emph{\bibinfo{journal}{Z. anorg. allg. Chem.}}
  \textbf{\bibinfo{volume}{539}}, \bibinfo{pages}{175} (\bibinfo{year}{1986}).

\bibitem{Muller:1991_a}
\bibinfo{author}{Muller, R.}, \bibinfo{author}{Kuckel, M.},
  \bibinfo{author}{Schuster, H.~U.}, \bibinfo{author}{Muller, P.} \&
  \bibinfo{author}{Bronger, W.}
\newblock \bibinfo{title}{Neue amnx-verbindungen mit a-Rb, Cs und x-P, As, Sb,
  Bi: Struktur und magnetismu}.
\newblock \emph{\bibinfo{journal}{J. Alloys and Compounds}}
  \textbf{\bibinfo{volume}{176}}, \bibinfo{pages}{167} (\bibinfo{year}{1991}).

\bibitem{Schucht:1999_a}
\bibinfo{author}{Schucht, F.} \emph{et~al.}
\newblock \bibinfo{title}{The magnetic properties of the alkali metal manganese
  pnictides KMnP, RbMnP, CsMnP, RbMnAs, KMnSb, KMnBi, RbMnBi, and CsMnBi -
  neutron diffraction and susceptibility measurements}.
\newblock \emph{\bibinfo{journal}{Z. anorg. allg. Chem.}}
  \textbf{\bibinfo{volume}{625}}, \bibinfo{pages}{31} (\bibinfo{year}{1999}).

\bibitem{Khmelevskyi:2008_a}
\bibinfo{author}{Khmelevskyi, S.} \& \bibinfo{author}{Mohn, P.}
\newblock \bibinfo{title}{Layered antiferromagnetism with high neel temperature
  in the intermetallic compound mn2au}.
\newblock \emph{\bibinfo{journal}{Appl. Phys. Lett.}}
  \textbf{\bibinfo{volume}{93}}, \bibinfo{pages}{162503}
  (\bibinfo{year}{2008}).

\bibitem{Scholl:2004_a}
\bibinfo{author}{Scholl, A.}, \bibinfo{author}{Liberati, M.},
  \bibinfo{author}{Arenholz, E.}, \bibinfo{author}{Ohldag, H.} \&
  \bibinfo{author}{{St\"{o}hr}, J.}
\newblock \bibinfo{title}{Creation of an antiferromagnetic exchange spring}.
\newblock \emph{\bibinfo{journal}{Phys. Rev. Lett.}}
  \textbf{\bibinfo{volume}{92}}, \bibinfo{pages}{247201}
  (\bibinfo{year}{2004}).

\end{thebibliography}

%

\section*{Corresponding author}
Correspondence and requests for materials should be addressed to Tomas Jungwirth, jungw@fzu.cz,  Institute of Physics ASCR, v.v.i., Cukrovarnick\'a 10, 162 53 Praha 6, Czech Republic.
\section*{Acknowledgment}
We thank  Petr Ji\v{r}\'{i}\v{c}ek, Zbyn\v{e}k \v{S}ob\'a\v{n}, and Miroslav Mary\v{s}ko for experimental assistance and we acknowledge support  from EU Grant FP7-214499 NAMASTE, FP7-215368 SemiSpinNet, from Czech Republic Grants AV0Z10100520, AV0Z10100521, MSM0021620834, MSM0021620857, KAN400100652, LC510, and Preamium Academiae. 
\section*{Author contributions}
MBE growth: VN, MC, TJ; experiments and data analysis: XM, VN, PH, PN, VH, JZ, PK, TJ; MBE end experimental conceptual and technical assistance: CTF, RPC, BLG, IN; theory: FM, ABS, JM, TJ; writing: TJ,VN.
\section*{Competing financial interests}
The authors declare that they have no competing financial interests.

\begin{figure}[t]
\hspace*{0cm}\epsfig{width=1\columnwidth,angle=0,file=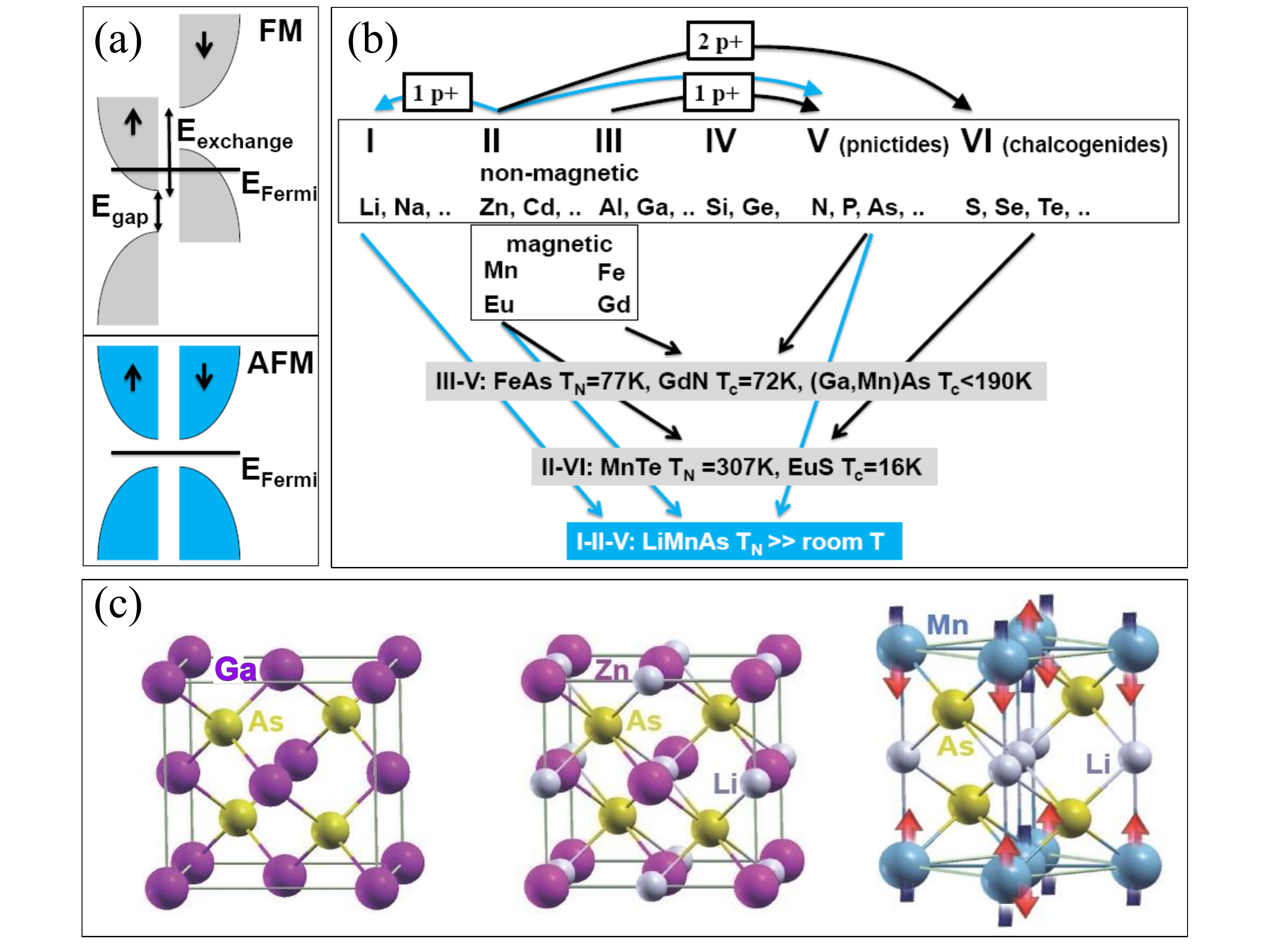}
\vspace*{-0.5cm}
\caption{(a) Illustration of the competition between the band-gap and the FM exchange-splitting which results in the Fermi energy inside the band, i.e. metal character, for large enough exchange-splittings. The absence of this competition in AFMs makes strong AFM semiconductors more likely to occur than strong FM semiconductors. (b) The illustration of the survey of magnetic semiconductor counterparts to conventional semiconductors with eight valence electrons per primitive cell. Consistent with panel (a), most of the magnetic semiconductors are AFMs and the more rare FM semiconductors are found primarily among compounds with more localized (less hybridized) $f$-electron magnetic elements and lighter anions (wider band-gaps). Magnetic transition temperatures safely above room temperature are found in the compounds containing the group-I element and I-Mn-V's are the simplest representatives of this class of materials. (c) Schematic 3D plots of the crystal structure of a conventional zinc-blende III-V semiconductor  (GaAs), of a non-magnetic I-II-V semiconductor (LiZnAs) and of the magnetic I-Mn-V compound (LiMnAs). The unit cells are chosen to highlight the similarities between the respective crystal structures.
}
\label{f1}
\end{figure}

\protect\newpage
\begin{figure}[t]
\hspace*{0cm}\epsfig{width=1\columnwidth,angle=0,file=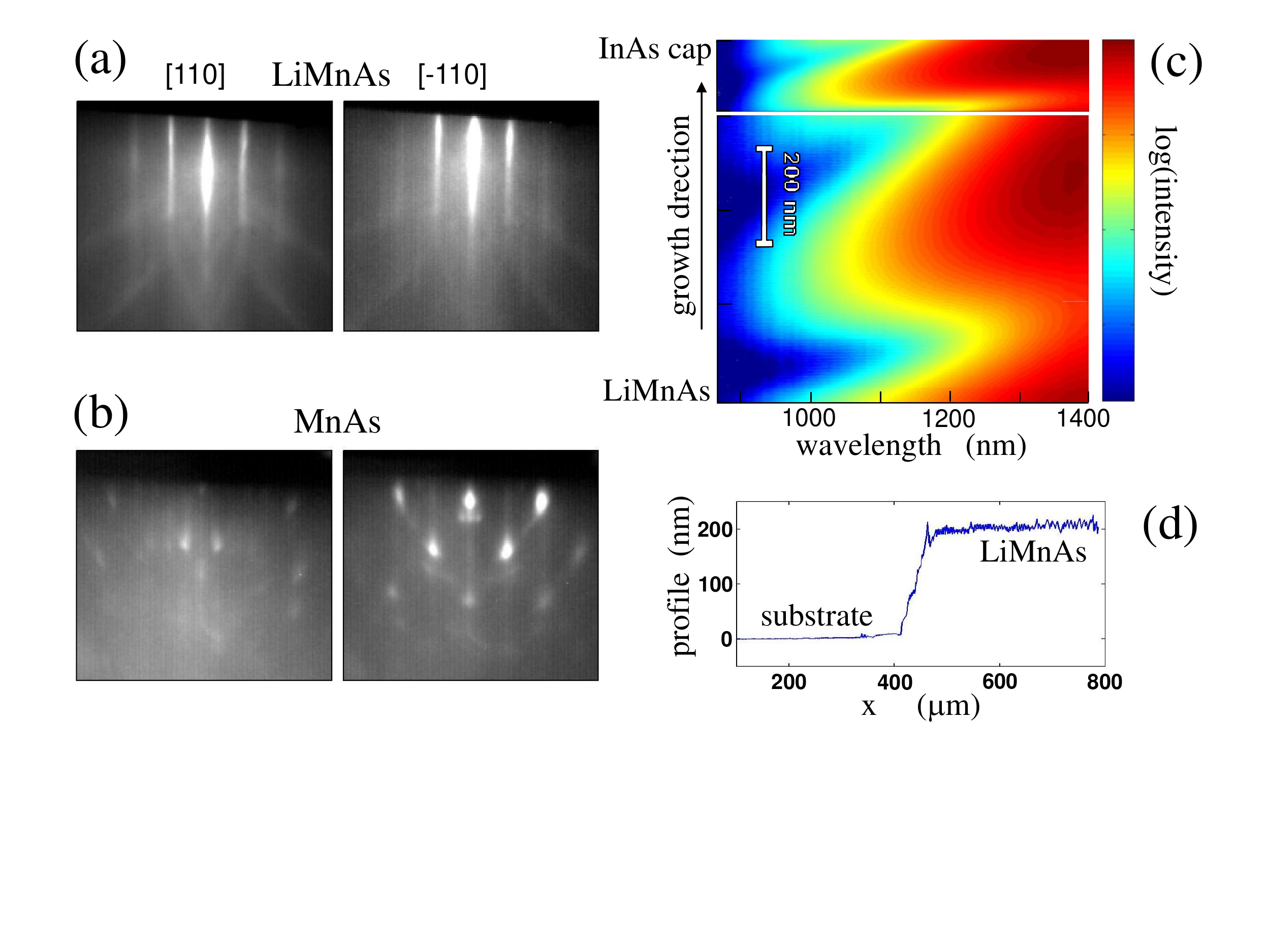}
\vspace*{-0.5cm}
\caption{RHEED images of the LiMnAs film (a) and of the MnAs film (b) after 60 minutes 
of MBE growth at identical conditions, except for the Li cell closed in case of the MnAs. (c) Fabry-P\'erot interference oscillations
of the light back-reflected by the growing LiMnAs film on an InAs substrate plotted as a function of the growth-time and wavelength of the detected light. The oscillatory interferences are typical of a semiconductor film. (d) Thickness profile of a LiMnAs wafer across the edges masked by the sample holder during the growth. The total growth time of this wafer was 60 minutes, resulting in $\sim 200$~nm thick LiMnAs epilayer. }
\label{f2}
\end{figure}

\protect\newpage
\begin{figure}[t]
\vspace*{-1.2cm}
\hspace*{0cm}\epsfig{width=0.9\columnwidth,angle=0,file=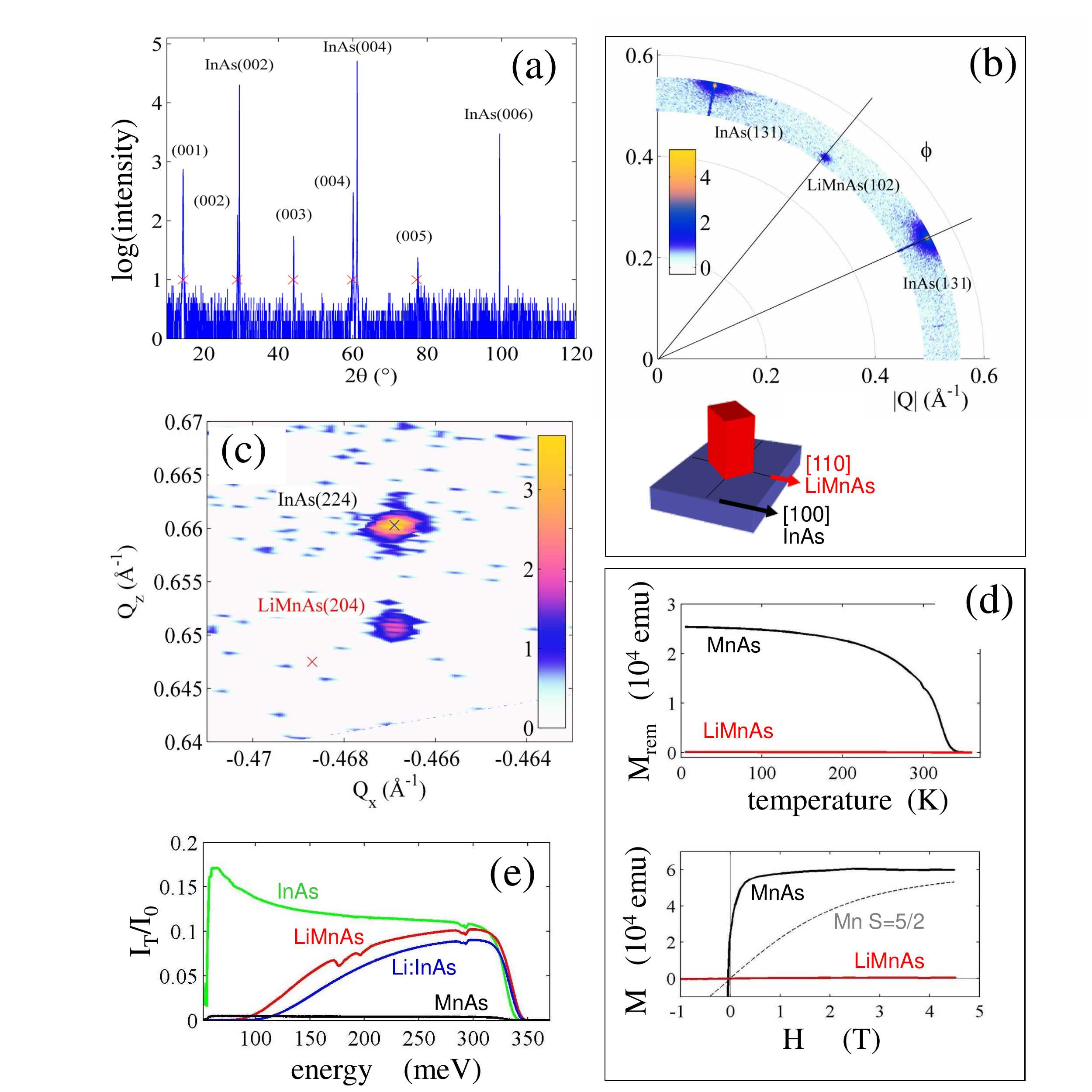}
\vspace*{-0cm}
\caption{(a) Full set of (001) oriented LiMnAs X-ray reflections. Bulk values are denoted by crosses. (b) Azimuthal scans as a function of the wavevector $Q$. The corresponding 45$^{\circ}$ in-plane rotation of the LiMnAs unit cell with respect to the InAs substrate is illustrated in the inset. (c) Reciprocal space maps evidencing the vertical alignment of the substrate peak and the peak of the strained LiMnAs film. The black and red crosses in the plot denote the expected positions for the substrate and bulk LiMnAs, respectively. (d) Temperature dependent remanence at 0.2~mT and magnetic field dependent magnetization at 4~K of a 200~nm thick LiMnAs epilayer (red curve) and of the FM MnAs sample  (solid black curve) containing the same amount of Mn. For comparison, we also show the theoretical Brillouin function which describes magnetization at 4~K of an equal number of Mn atoms represented by uncoupled paramagnetic $S=5/2$ spins in the external magnetic field.  (e) Transmissivity of the LiMnAs/InAs sample (red curve) in the near infra-red region measured {\em ex situ} and 
compared to the MnAs/InAs sample (black curve), to an unprocessed InAs substrate (green curve), 
and to a Li-doped InAs substrate (blue curve). }
\label{f3}
\end{figure}

\protect\newpage
\begin{figure}[t]
\hspace*{0cm}\epsfig{width=1\columnwidth,angle=0,file=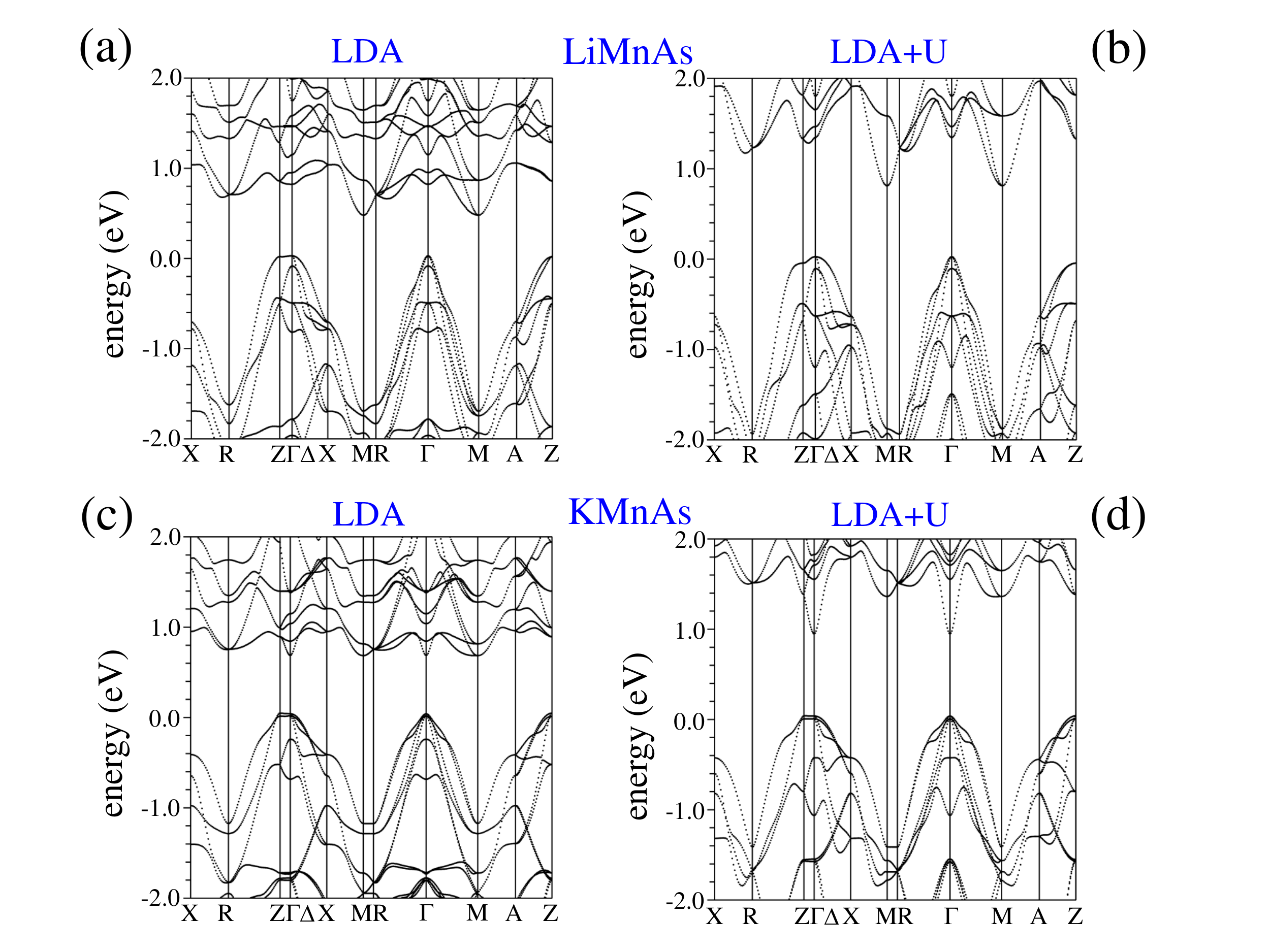}
\vspace*{0.5cm}
\caption{(a) Full-potential LDA calculations of the band dispersions of AFM LiMnAs. Spin-orbit coupling is turned off in this plot for clarity.  (b) LDA+U band dispersions of LiMnAs. (c) LDA, (d) LDA+U band dispersions of KMnAs.}
\label{f4}
\end{figure}

\protect\newpage
\begin{figure}[t]
\hspace*{0cm}\epsfig{width=1\columnwidth,angle=0,file=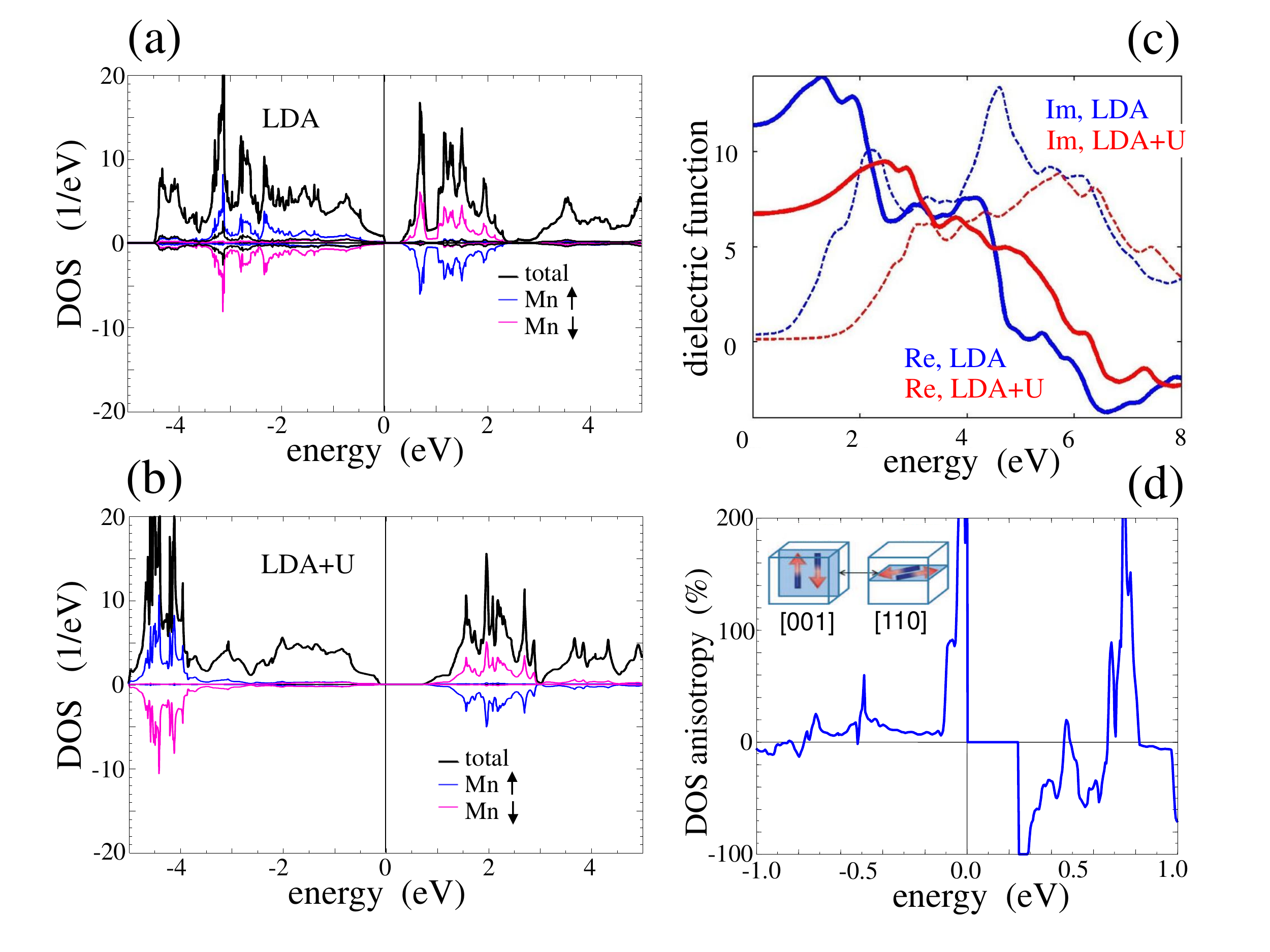}
\vspace*{0.5cm}
\caption{Total and element-resolved density of states of LiMnAs calculated in (a) LDA and (b) LDA+U. (c) Complex dielectric function of LiMnAs in the out-of-plane direction calculated in the LDA and LDA+U. (d) Relativistic full-potential LDA calculations of the anisotropy in the density of states with respect to the staggered moment orientation along the [001] and [110] crystal axes. These anisotropies are order of magnitude larger than in the metal Mn-based AFMs and vary strongly near the valence band and conduction band edges.
}
\label{f5}
\end{figure}
\end{document}